\documentclass{article}

\usepackage{algorithm2e}
\usepackage{boxedminipage}

\usepackage{amsmath, amsthm, amssymb, bbm, bm, esvect}
\usepackage{graphicx}
\usepackage{verbatim}
\usepackage{natbib}
\usepackage{caption}
\usepackage{subcaption}
\usepackage{fancyvrb}
\usepackage{enumerate}
\usepackage{relsize}
\usepackage{scalerel}
\usepackage{varioref}

\usepackage{array,multirow,graphicx}
\usepackage{ragged2e}
\usepackage{float}

\usepackage{tikz}
\usetikzlibrary{fit,positioning}

\usepackage{hyperref}
\usepackage[margin=1.5in]{geometry}
\usepackage{chngpage}
\hypersetup{colorlinks,citecolor=blue,urlcolor=blue,linkcolor=blue}

\usepackage{stefan_tex}
\graphicspath{{./figures/}}

\theoremstyle{plain}
\newtheorem{prop}{Proposition}

\newtheorem{theo}[prop]{Theorem}

\theoremstyle{definition}

\newtheorem{assu}{Assumption}

\theoremstyle{remark}

\newtheorem{rema}{Remark}

\date{Draft version \ifcase\month\or
	January\or February\or March\or April\or May\or June\or
	July\or August\or September\or October\or November\or December\fi \ \number%
	\year\ \  }

\author{
	Xinkun Nie \\ \texttt{xinkun@stanford.edu}\and 
	Guido Imbens \\ \texttt{imbens@stanford.edu} 
	\and
	Stefan Wager \\ \texttt{swager@stanford.edu}}

\title{Covariate Balancing Sensitivity Analysis for \\ Extrapolating Randomized Trials across Locations}

\begin{document}

\maketitle
\begin{abstract}
The ability to generalize experimental results from randomized control trials (RCTs) across locations is crucial for informing policy decisions in targeted regions. Such generalization is often hindered by the lack of identifiability due to unmeasured effect modifiers that compromise direct transport of treatment effect estimates from one location to another. We build upon sensitivity analysis in observational studies and propose an optimization procedure that allows us to get bounds on the treatment effects in targeted regions. Furthermore, we construct more informative bounds by balancing on the moments of covariates. In simulation experiments, we show that the covariate balancing approach is promising in getting sharper identification intervals.	
\end{abstract}

\section{Introduction}

Randomized control trials (RCTs) have long been the gold standard in evaluating public policy. In order to leverage insights learned from RCTs conducted in one location to drive policy change in other locations, there remains the concern about external validity, namely that the target population affected by the considered policy change may respond to the intervention differently than the population sample included in the RCT study \citep{cook2002experimental, olsen2013external}. For example, perhaps the randomized study was run in Texas, but we want to use results from the study to inform policy across all 50 states. Or perhaps participation in the randomized study was voluntary. As a concrete example, \cite{attanasio2011subsidizing} examine the effect of vocational training on labor market outcomes using a randomized study—but participants needed to apply to be part of the study, and so we may be concerned that the effect of vocational training for study participants may differ from the effect of training on the general population.

In cases where any cross-location divergences in populations can be explained away using observed covariates, the generalization problem is addressed by several authors, including \citet{dehejia2019local, hirshberg2019minimax, hotz2005predicting,   stuart2011use}, using ideas that generalize propensity score methods going back to \citet{rosenbaum1983central}. In this paper, however, we are most interested in the case where locations differ along unobserved features, and so controlling for observed covariates doesn’t solve the cross-location generalization problem.

As discussed further below, our proposal builds upon the literature on sensitivity analysis in observational studies \citep{rosenbaum2002design}, and in particular on recently proposed methods for sensitivity analysis via linear programming \citep{aronow2012interval, miratrix2017shape, zhao2019sensitivity}. This literature is focused on cases where a randomized or observational study is confounded by unobserved features, and so the resulting treatment effect estimates do not have internal validity. Here, in contrast, we are not worried about the internal validity of our study; rather, we are concerned about how unobserved features may affect external validity when we aim to extrapolate results from RCT studies to other locations.

Our main finding is that in the cross-study generalization problem we have access to richer
information than in the single-study sensitivity analysis problem, and that we can use this information to substantially improve power. At a high level, our approach is driven by the insight that we can collect data on observable features in both the location where the RCT was conducted and the target location we wish to extrapolate to, and that any selection weights that transport treatment effect estimates from one location to the other must respect moments of these observed covariates. As such, our motivation is qualitatively related to recent work on covariate-balancing estimation in causal inference \citep{athey2018approximate, graham2016efficient, hainmueller2012entropy, imai2014covariate, zubizarreta2015stable}.

\subsection{Related Work}
Generalizing results from randomized trials run in one location to other locations has been a central question in economic and healthcare policymaking \citep[e.g.,][]{ cole2010generalizing, gechter2015generalizing, hotz2005predicting}. Several authors have discussed generalizability bias (or transportability bias) that arises when there is either selection bias in a RCT or we try to generalize the results of one RCT to a different location that has a different covariate distribution \citep{bareinboim2016causal,  olsen2013external, pressler2013use}. To address covariate shift in the case of no measured confounding, many existing works use the generalized propensity scores for either matching or weighting \citep{cole2010generalizing,  dehejia2019local, hirshberg2017augmented,stuart2011use, westreich2017transportability}, or covariate matching \citep[e.g.,][]{gretton2009covariate}. We also note the recent advances in the literature on transportability in causal graphical models that focuses on designing graphical algorithms for identifying whether transportability is feasible from graphs \citep[e.g.,][]{bareinboim2013meta,  bareinboim2014transportability, bareinboim2016causal, pearl2014external}. Our work is complementary to theirs in the sense that we don't assume transportability is necessarily feasible due to unmeasured effect modifiers, and instead we compute identification intervals whose length varies with the distributional imbalance of these unobserved factors across locations.

We build upon the literature on sensitivity analysis in observational studies \citep{cornfield1959smoking, fogarty2016sensitivity, imbens2003sensitivity, rosenbaum1987sensitivity, rosenbaum2002attributing, rosenbaum2002design, rosenbaum2010design, rosenbaum2011new, rosenbaum2014weighted,  shen2011sensitivity,  vanderweele2017sensitivity} and in particular, the recently developed works on using linear programming \citep{aronow2012interval, miratrix2017shape,yadlowsky2018bounds,  zhao2019sensitivity}. This line of work focuses on using sensitivity analysis to address internal validty of an observational study. On the other hand, our work is focused on addressing external validty of an RCT. The sensitivity model we have employed is closest in connection with the marginal sensitivity model advocated in \citet{tan2006distributional} and \citet{zhao2019sensitivity}.

Given that violations of transport assumptions is untestable for generalizing results from RCT to other locations, a few authors have proposed methods that perform a sensitivity analysis on how much violation on this assumption can lead to generalizability bias.  \cite{andrews2017weighting} considers the limit of the role of private information, and their approach only applies to the special case where the randomized study population is a subset of the target population we wish to apply a policy intervention to.  \cite{gechter2015generalizing} considers the range of distributions of the treated potential outcomes conditioning on the controls potential outcomes. 
 \cite{nguyen2017sensitivity} 
considers the ranges on the mean of the unobservabed covariates.  We view these approach as complementary to our proposal, as we consider the limit on the ratio of the probability density function on the unobserved covariates among locations, and further exploit covariate balance for improved power.

The core of our proposal is that by leveraging covariate information in the locations of interest, the set of weights for weighting outcomes can be used to balance moments on the covariates between locations. This would shorten identification intervals for the target population compared to prior work. In particular, the idea of covariate balancing has become popular in other contexts such as estimating average treatment effect \citep{athey2018approximate,     bennett2018building, graham2016efficient,hainmueller2012entropy, hirshberg2017augmented, kallus2018balanced, wang2017minimal, zhao2019covariate,zubizarreta2015stable}. 

We note that there is a growing interest in combining observational data with RCTs \citep{athey2016estimating, kaizar2011estimating, kallus2018removing, rosenman2018propensity} to improve power. In our work, we focus on only using the RCT data, and leave it to future work on how to incorporate observational data in our framework. We also note the interesting and promising direction of using Bayesian hierarchical models for combining RCT results from multiple studies   \citep[e.g.,][]{meager2016aggregating,vivalt2016much}. They work in the setting of using aggregated experimental data on the study level, whereas our work leverages individual-level data. There is also a growing literature on transfer learning for domain shifts (see, e.g., \citet{bastani2020predicting} and references therein for a review); in our case, we focus on transfer results from a randomized trial from one location to another.

\section{Extrapolating on Observables and Unobservables}
\subsection{Formal Model and Assumptions}
We assume there are two locations (e.g., two cities). In the first location, an RCT has been conducted to evaluate the impact of an intervention on some outcome of interest, e.g., whether a job training program improved participants' earnings. In the second location, without running an additional RCT tailored for this new population, we want to ask the question of to what extend the previous RCT result is applicable to the new location. In both locations, we observe a set of covariates such as each citizen's age and marital status, but we might not have access to other important covariates such as their education level. 

Formally, we denote the two locations by Location $0$ and Location $1$. Location $0$ is where the RCT is conducted, and we wish to generalize the RCT results to Location $1$. Throughout the paper, we use the subscripts of $0$ and $1$ to denote respective quantities for the two locations. We assume the data are i.i.d generated in each location: $X_i, U_i, W_i, Y_i(0), Y_i(1) \overset{\text{i.i.d}}{\sim}  P_k$ for $i=1, \cdots, n_k$ where the subscript $k\in \{0,1\}$ corresponds to the two locations respectively, $X \in \mathcal{X}$ is the observed covariates, $U \in \mathcal{U}$ is the unmeasured covariates, $W$ is the randomized treatment assignment indicator where $W=1$ indicates that the treatment is assigned, and $W=0$ indicates otherwise, and $Y(w)$ for $w=0,1$ is the potential outcome corresponding to having received treatment or lack thereof. In Location $0$, we observe data $X, Y, W$ with $Y=Y(W)$, and the propensity score $e:=P_0(W=1)$ is a known constant. On the other hand, in Location $1$, we assume we only observe the covariates $X$.  The causal estimand is $\tau = \EE[1]{Y(1)-Y(0)}$, denoting the average treatment effects in the target Location $1$. For convenience, we use $L_i \in \{0,1\}$ to denote the location indicator for each unit. 

We start with a few assumptions.

\begin{assu}[Conditional location independence]
	\label{assu:relevant-select}
	For all $x\in \mathcal{X}$ and for all $u\in\mathcal{U}$,
	$P_0[Y(1), Y(0) \cond X, U] =P_1[Y(1), Y(0) \cond X, U]$.
\end{assu}

This assumption implies that the distribution of the potential outcomes conditioning on the observed variables $X$ and any unmeasured effect modifiers $U$ is the same across the two locations. Most existing works in the literature assume transportability only conditioning on the observed covariates \citep[e.g.,][]{hotz2005predicting}. Given that the unmeasured effect modifier $U$ can have any association with the potential outcomes $Y(0)$ and $Y(1)$ conditionally on $X$, we note that Assumption \ref{assu:relevant-select} does not impose any meaningful restrictions on its own.

\begin{assu}[Support inclusion]
	\label{assu:support}
	$supp(P_1(X, U)) \subseteq supp(P_0(X, U))$. In particular, $P_1(X, U)$ is absolutely continuous with respect to $P_0(X, U)$, where the covariate density ratio $r(X, U) = \frac{dP_1(X, U)}{dP_0(X, U)}$ and the observed covariate density ratio $r(X) = \frac{dP_1(X)}{dP_0(X)}$ exist.
\end{assu}

\begin{prop}
	\label{prop:1}
Under Assumptions \ref{assu:relevant-select} and \ref{assu:support}, we have
\begin{align}
	\label{eq:tau_new_defn}
	\tau = \EE[1]{Y(1) - Y(0)} = \EE[0]{r(X,U)\EE[0]{Y(1)-Y(0)\cond X, U}}.
\end{align}
\end{prop}

The above proposition shows that if the unmeasured effect modifiers were known and measured in both locations, then the causal quantity of interest $\tau$ is  identified. In order to turn the above into an estimator,  we could use a standard inverse weighted estimator:
\begin{align}
	\label{eq:unbiased-ipw}
	\hat{\tau}^*
	&= \frac{1}{n_0}\sum_{i: L_i=0} \frac{r(X_i, U_i)W_i  Y_i}{P(W_i=1)} - \frac{1}{n_0}\sum_{i: L_i=0} \frac{r(X_i, U_i) (1-W_i) Y_i}{P(W_i=0)}.
\end{align}
such that $\EE[0]{\hat{\tau}^*} = \tau$.
Since $U$ is unobserved, the above estimator is not feasible.  Instead, we further allow sensitivity models detailed in the next subsections which assume bounds on the ratio of the conditional probability density of unmeansured effect modifiers $U$. This would in turn allow us to derive bounds for $\hat{\tau}$ via a linear programming optimization. 

\section{A sensitivity model for domain extrapolation}
 To relate $r(x,u)$ and $r(x)$, we define the unobserved distribution shift ratio as
 \begin{align*}
 	z^*(x,u) := \frac{r(x,u)}{r(x)}
 \end{align*}
and use the shorthand $z_i^*:= z^*(X_i, U_i)$. These $z_i^*$ capture the amount by which the unobserved
effect modifiers $U_i$ affect the oracle estimator \eqref{eq:unbiased-ipw}, which we can re-write as
\begin{align}
	\label{eq:unbiased-ipw-z}
	\hat{\tau}^*
	&= \frac{1}{n_0}\sum_{i: L_i=0} \frac{z^*_i r(X_i) W_i  Y_i}{P(W_i=1)} - \frac{1}{n_0}\sum_{i: L_i=0} \frac{z^*_i r(X_i) (1-W_i) Y_i}{P(W_i=0)}.
\end{align}
 Given we don't know the density ratio terms $r(X, U)$ in \eqref{eq:unbiased-ipw} due to the unmeasured effect modifier $U$, we instead aim to get bounds on $\tau$ by estimating $r(X)$ which does not depend on $U$ and by assuming a bound on $z^*(x,u)$.  In particular, we assume the following sensitivity model that directly implies a bound on $z(x,u)$. 
 
\begin{assu}[Transport sensitivity model]
	\label{assu:marginal-sensi-transport}
	There exists $\Gamma^* \geq 1$ such that ${\Gamma^*}^{-1} \leq P_1(U=u \cond X=x) / P_0(U=u \cond X=x) \leq \Gamma^*$ for all $x \in \mathcal{X}$ and all $u \in \mathcal{U}$.
\end{assu}

By Bayes rule, we immediately have the following:
\begin{prop}
\label{prop:z}
By Assumption \ref{assu:marginal-sensi-transport}, $z^*(x,u) \in [{\Gamma^*}^{-1}, \Gamma^*]$ for all $x\in \mathcal{X}$ and $u \in \mathcal{U}$.
\end{prop}
Analogous sensitivity models are common in the sensitivty analysis literature for observational studies \citep[e.g.][]{rosenbaum2002design, zhao2019sensitivity} to assess robustness of findings to unmeasured confounding.



\section{Bounds via linear programming}

With a bound on $z^*(x,u)$ and an estimated density ratio $\hat{r}(\cdot)$, we can derive bounds on $\hat{\tau}$ from \eqref{eq:unbiased-ipw-z}.
Since we don't know the ground truth value of $\Gamma^*$, we supply a $\Gamma$ value and solve the following optimization problem to get the upper bound $\tilde{\tau}^+$. 
\begin{align}
	\label{eq:opt-no-bal-any-propensity}
	\tilde{\tau}^+ &= \sup \frac{\sum_{i: L_i=0} W_i Y_i \hat{r}_i(X_i) z_i}{\sum_{i: L_i=0} W_i }\nonumber\\
	&\ \ \ \ \ \ \ \  - \frac{\sum_{i: L_i=0} (1-W_i)Y_i \hat{r}(X_i)z_i}{\sum_{i: L_i=0} (1-W_i)} \nonumber\\
	&\ \ \ \ s.t. \ \Gamma^{-1} \leq z_i \leq \Gamma, 1\leq i \leq n_0, 
\end{align}
To get the lower bound $\tilde{\tau}^-$, we instead take the infimium in the optimization above. This key idea of employing a linear program to bound the target estimand when the density ratio is unknown but can vary within some range is also seen in previous literature. For example, \citet{aronow2012interval} and \citet{miratrix2017shape} employ similar ideas for identification of a population mean when the sampling selection weight is unknown.
\begin{rema}
In practice, we suggest varying $\Gamma$ across a wide range to report the upper and lower bounds $\tilde{\tau}^{-}$ with domain experts guiding on realistic choices of $\Gamma$. We also note that in the context of potential violations of exogeneity in observational studies,  \citet{imbens2003sensitivity} proposes leveraging observed covariates to judge the plausbility of sensitivity parameters. We leave it to future work to investigate similar heuristics for guiding choices on $\Gamma$.
\end{rema}

The roadmap for the rest of this section is as follows: We first show how to estimate $\hat{r}$ in a way that allows us to achieve sharper bounds. Next, we use the estimate $\hat{r}$ to get the upper bound by \eqref{eq:opt-no-bal-any-propensity} (and respectively, the lower bound by taking the infimum of the same optization) for $\tilde{\tau}$. Finally, we conclude with showing the bounds $[\tilde{\tau}^-, \tilde{\tau}^+]$ gives consistent coverage for the ground truth treatment effect $\tau$ in Location 1. 

First, we estimate $\hat{r}(\cdot)$. While \eqref{eq:opt-no-bal-any-propensity} takes in $\hat{r}(\cdot)$ from any density ratio estimators, we suggest 
estimating $r(x)$ directly by moments matching \citep{	bradic2019sparsity, imai2014covariate, ning2017high, qin1998inferences, sugiyama2012density}.  Doing so would allow us to effectively shorten the estimation bounds as detailed later in Section \ref{sec:shorten}. To fix ideas, we denote $g(x) = \log r(x)$ and then assume the following:

\begin{assu}
	\label{assu:logistic}
	We have access to basis function $\phi(\cdot) \in \RR^p$ such that $\Norm{\phi(\cdot)}_\infty \in [-M, M]$ for some $M>0$,
and	$g(x) = \phi(x) \cdot \beta_\phi$ for some $\beta_\phi \in \RR^p$.
\end{assu}
When the basis function $\phi(\cdot)$ is the identity, this simply implies that the density ratio follows a logistic form. On the other hand, we can take $\phi(\cdot)$ to be any bounded basis expansion, which makes the above assumption not as restrictive. In Section \ref{sec:approx}, we further relax this assumption and incorporate the model misspecification error. For the rest of the paper, we assume $\phi(X_i)$ contains an intercept term.

By Assumption \ref{assu:support} and the fact that an RCT is conducted in Location $0$, 
we have the following covariates balancing moment condition via a change of measure: for $w=0,1$,
\begin{align}
	\label{eq:balance-x}
	\EE[0]{ \phi(X_i)e^{g(X_i)}  \cond W_i=w} = \EE[1]{\phi(X_i)}
\end{align}
We exploit the empirical counterpart of the above moment equation to estimate the  $l(x)$.  Given Assumption \ref{assu:logistic}, we only need to estimate $\beta^{(w)}$.  Concretely, we solve the following optimization problem:
\begin{align}
	\label{eq:learn-beta}
	\hat{\beta}_\phi^{(w)} = \argmin_{\beta} \frac{ \sum_{i: L_i=0} \mathbbm{1}_{W_i=w} \exp({\phi(X_i)}^\top \beta)}{\sum_{i: L_i=0}  \mathbbm{1}_{W_i=w}}  - \frac{\sum_{i: L_i=1} \phi(X_i)^\top \beta}{n_1},
\end{align}
and then set
\begin{align}
	\label{eq:learn-g}
	\hat{g}_\phi(X_i) = \sum_{w=0}^1\mathbbm{1}_{W_i=w} \phi(X_i)^\top \hat{\beta}_\phi^{(w)}.
\end{align}
The reason we fit \smash{$\hat{\beta}_\phi^{(w)}$} separately on the treated and control units is that it enables us to exactly match the empirical
version of the moment condition in \eqref{eq:balance-x} for both treated and control groups
(this follows immediately form the first-order condition in \eqref{eq:learn-beta}):
\begin{equation}
\frac{ \sum_{i: L_i=0} \mathbbm{1}_{W_i=w} \phi(X_i) \hat{g}_\phi(X_i) }{\sum_{i: L_i=0}  \mathbbm{1}_{W_i=w}} = \frac{\sum_{i: L_i=1} \phi(X_i)^\top \beta}{n_1}, \  \eqfor \  w \in 0, \, 1.
\end{equation}
We note that, if \smash{$\beta_\phi$} is unique, then both \smash{$\hat{\beta}_\phi^{(w)}$}
 will eventually converge to \smash{$\beta_\phi$} in large samples.

By Assumption \ref{assu:logistic}, the density ratio is identified, and the estimated density ratio is consistent, i.e. $\Norm{\hat{g}_\phi(x) - g(x)}_\infty \to_p 0$ \citep{qin1998inferences, sugiyama2012density}. There is an exact mapping between estimating density ratios and estimating location propensity by treatment locations as random variables. We use a covariate balancing estimator for the density ratio, which corresponds to the covariate balancing estimator for propensity scores if we treated the locations as random variables. In the line of work for propensity score estimation, our covariate balancing approach is closely related to \citet{imai2014covariate, tan2017regularized, tan2018model,zhao2019covariate}. We can then substitute in $\exp\p{\hat{g}_\phi(\cdot)}$ for $\hat{r}(\cdot)$ in \eqref{eq:opt-no-bal-any-propensity} to compute the corresponding upper and lower bounds $\tilde{\tau}^+$ and $\tilde{\tau}^-$. 

Next, we show the resulting bounds give consistent coverage:
\begin{theo}
	\label{thm:consistency-1}
	 Under Assumptions \ref{assu:relevant-select}, \ref{assu:support}, \ref{assu:marginal-sensi-transport} and \ref{assu:logistic} with $\Norm{\hat{r}(\cdot)-r(\cdot) }_\infty \to_p 0$, then with any $\Gamma \geq \Gamma^*$ in \eqref{eq:opt-no-bal-any-propensity}, for any $\epsilon > 0$, 
	\begin{align}
		\lim_{n_0\to \infty}\PP{\tau \in (\tilde{\tau}^- - \epsilon, \tilde{\tau}^+ + \epsilon)} \to 1
	\end{align}
	where $\tau =  \EE[1]{Y(1) - Y(0)}$.
\end{theo}

\section{Shorter bounds via covariate balancing}
\label{sec:shorten}
Although the optimization formulation in \eqref{eq:opt-no-bal-any-propensity} gives consistent coverage on $\tau$, the resulting bounds can often be too wide in practice to provide meaningful conclusions about the treatment effect $\tau$. As a concrete example, assume the age distribution is the same for the location that we have conducted the RCT in and the location we wish to extrapolate findings to. We may find that from the RCT, the treatment effect is largest among the young population. The optimization procedure detailed in the previous section does not leverage information on age distributions in these two locations, and could construct weights $z_i$ that overweight or underweight the young population leading to wide bounds.  

The key contribution of this work is to sharpen the estimation bounds by further leveraging covariate information available in both locations. In particular, similar to \eqref{eq:balance-x}, we can also balance the following moments using a change of measure that conditions on both the observed $X$ and the unmeasured effect modifiers $U$. Let $g(x,u) = \log r(x,u)$. Then the following moment equation holds: for $w=0,1$,
\begin{align}
	\label{eq:balance-xu}
	\EE[0]{ e^{g(X_i, U_i)}\phi(X_i) \cond W_i=w} = \EE[1]{\phi(X_i)}.
\end{align}
We then add the empirical counterpart of the moment equality constraints in \eqref{eq:balance-xu} to \eqref{eq:opt-no-bal-any-propensity} as the last two equalities (a) and (b) below.
\begin{align}
	\label{eq:bal-opt}
	\hat{\tau}^+ &= \sup \frac{\sum_{i: L_i=0} W_i z_i e^{\hat{g}_\phi(X_i)}Y_i}{\sum_{i: L_i=0} W_i}\nonumber\\
	&\ \ \ \ \ \ \ \  - \frac{\sum_{i: L_i=0} (1-W_i) z_i e^{\hat{g}_\phi(X_i)}Y_i}{\sum_{i: L_i=0} (1-W_i) } \nonumber\\
	&\ \ \ \ s.t. \ \Gamma^{-1} \leq z_i \leq \Gamma, 1\leq i \leq n_0,\nonumber \\
	&\ \ \ \ \ \ \ \ \ \frac{\sum_{i: L_i=0} W_i z_i e^{\hat{g}_\phi(X_i)}\phi(X_i)}{\sum_{i: L_i=0} W_i } \stackrel{(a)}{=} \frac{\sum_{i: L_i=1}^n  \phi(X_i)}{n_1} \nonumber\\
	&\ \ \ \ \ \ \ \ \ \frac{\sum_{i: L_i=0} (1-W_i) z_i e^{\hat{g}_\phi(X_i)}\phi(X_i)}{\sum_{i: L_i=0} (1-W_i) } \stackrel{(b)}{=} \frac{\sum_{i: L_i=1} \phi(X_i)}{n_1}.  \nonumber\\
\end{align}
Similarly, we can estimate $\hat{\tau}^-$ by taking infimum of the above optimization with the same set of constraints. Compared to \eqref{eq:opt-no-bal-any-propensity}, \eqref{eq:bal-opt} makes a few modifications. First, it substitutes in the estimate $\exp\p{\hat{g}_{\phi}(\cdot)}$ for $\hat{r}(\cdot)$ using \eqref{eq:learn-g}. Second, it adds two equality constraints (a) and (b). By setting the derivative of (a) and (b) to 0 in expectation, we arrive at the moment condition in \eqref{eq:balance-xu} for $w=0,1$ respectively.

Covariate balancing via \eqref{eq:bal-opt} shortens estimation intervals compared to solving \eqref{eq:opt-no-bal-any-propensity} due to the added equality constraints which limit the plausible range of $z_i$'s. To provide more intuition, we include in the Appendix a more technical comparison in the simple context where the covariates $X$ are discrete and the link function $\phi$ is the identity function. 

The constructed intervals from the above optimization \eqref{eq:bal-opt} gives consistent coverage of the underlying treatment effect parameter of interest.

\begin{theo}
	\label{thm:consistency}
	Consider a sequence of data generating processes for both of the two locations with population size $\{n_0, n_1\}$ such that $n_0, n_1\to\infty$ and $n_0/n_1 \to \eta$ for some constant $\eta>0$. Under Assumptions \ref{assu:relevant-select}, \ref{assu:support}, \ref{assu:marginal-sensi-transport} and \ref{assu:logistic} with $\hat{r}(\cdot) = \exp\p{\hat{g}_\phi(\cdot)}$ and $\hat{g}_\phi(\cdot)$ estimated with covariate balancing from \eqref{eq:learn-g}, then with any $\Gamma \geq \Gamma^*$ in \eqref{eq:bal-opt}, for any $\epsilon > 0$, 
	\begin{align}
		\lim_{\substack{n_0, n_1\to\infty \\ n_0/n_1\to\eta}}\PP{\tau \in (\hat{\tau}^- - \epsilon, \hat{\tau}^+ + \epsilon)} \to 1
	\end{align}
 where $\tau =  \EE[1]{Y(1) - Y(0)}$.
\end{theo}
This implies that as the sample size increases, if we employ a $\Gamma$ value in the optimization that is no less than what is needed for Assumption \ref{assu:marginal-sensi-transport} to hold, then the constructed interval from the balancing estimator eventually covers the true treatment effect in the target location that we wish to apply policy interventions to.

\begin{rema}
	Because we learn the density ratio $\hat{r}(x)$ via covariate balancing, the optimization \eqref{eq:bal-opt} is always feasible by taking $z_i=1$. This is a key proof component for Theorem \ref{thm:consistency}, as Slater's conditions would hold, which allows us to leverage the zero duality gap to complete the proof.
\end{rema}\textit{}

\section{Bounds under model misspecification}
\label{sec:approx}
So far, we have assumed that the density ratio follows a logistic model with a basis expansion $\phi(\cdot)$ by Assumption \ref{assu:logistic}. In this section, we relax this assumption and develop bounds that take model misspecification error into account.

Given any basis function $\phi(\cdot)$, let $\beta_\phi^{(w)}$ be the population minimizer, i.e. for $w=0,1$,
\begin{align}
	\beta_\phi^{(w)} = \argmin_{\beta}\EE[0]{ e^{\beta^\top \phi(X_i)} \phi(X_i) \cond W_i=w} - \beta^\top\EE[1]{\phi(X_i)},
\end{align} 
and let
\begin{align}
	g_\phi(X_i) = \sum_{w=0}^1\mathbbm{1}_{W_i=w} \phi(X_i)^\top \beta_\phi^{(w)}
\end{align}
be the logistic approximation to the ground truth density ratio logit function $g(\cdot)$. For the rest of this section, instead of Assumption \ref{assu:marginal-sensi-transport}, we build upon the following sensitivity model instead.
\begin{assu}[Sensitivity model with model misspecification]
	\label{assu:marginal-sensi-transport-approx}
	For a given basis function $\phi(\cdot)$, ${\Gamma^*}^{-1}  e^{g_\phi(x)}/e^{g(x)} \leq P_1(U=u \cond X=x) / P_0(U=u \cond X=x) \leq \Gamma^* e^{g(x)}/ e^{g_\phi(x)}$ for all $x \in \mathcal{X}$ and all $u \in \mathcal{U}$.
\end{assu}
Then immediately by Bayes rule, we have
\begin{prop}
	\label{prop:z-misspe}
	By Assumption \ref{assu:marginal-sensi-transport-approx}, $z^*(x,u) \in [{\Gamma^*}^{-1}e^{g_\phi(x)}/e^{g(x)} , \Gamma^*e^{g(x)}/ e^{g_\phi(x)}]$ for all $x\in \mathcal{X}$ and $u \in \mathcal{U}$.
\end{prop}

In practice, we don't know the magnitude of $e^{g(x)} / e^{g_\phi(x)}$. We supply an additional sensitivity parameter $M$ such that $M \geq e^{g(\cdot)} / e^{g_\phi(\cdot)}$ based on how much we believe the density ratio is misspecified with a chosen logistic model. We proceed to get bounds on $\tau$ by estimating $\hat{g}_\phi(\cdot)$ via \eqref{eq:learn-g}. We solve for $\hat{\tau}_\phi^+$ by \eqref{eq:bal-opt} but we substitute in $\Gamma M$ for $\Gamma$, and similarly we take the infimum to derive $\hat{\tau}_\phi^-$. Compared to the previous sections, we relax the constraint that $g(x)$ is well specified with a logistic form, and we conclude with the same consistency result if we assume Assumption \ref{assu:marginal-sensi-transport-approx} instead of Assumption \ref{assu:marginal-sensi-transport}.
\begin{theo}[Consistency under model misspecifcation]
	\label{thm:consistency-approx}
Consider a sequence of data generating processes for both of the two locations with population size $\{n_0, n_1\}$ such that $n_0, n_1\to\infty$ and $n_0/n_1 \to \eta$ for some constant $\eta>0$. 	Assume for some basis expansion $\phi(\cdot)$, $\hat{g}_\phi(\cdot)$ is estimated with covariate balancing from \eqref{eq:learn-g}, and that Assumptions \ref{assu:relevant-select}, \ref{assu:support}, and \ref{assu:marginal-sensi-transport-approx} hold with sensitivity parameter $\Gamma^*$, then with any $\Gamma \geq \Gamma^*$ and $M \geq e^{g(\cdot)} / e^{g_\phi(\cdot)}$ in \eqref{eq:bal-opt-approx}, for any $\epsilon > 0$,
	\begin{align*}
		\lim_{\substack{n_0, n_1\to\infty \\ n_0/n_1\to\eta}}\PP{\tau \in (\hat{\tau}_\phi^--\epsilon, \hat{\tau}_\phi^+ +\epsilon)} \to 1.
	\end{align*}
	where $\tau =  \EE[1]{Y(1) - Y(0)}$.
\end{theo}

\section{California GAIN Program}
We apply our proposed estimator to the California Greater Avenue for Independence (GAIN) dataset \citep{hotz2006evaluating}. A policy analyst may run an RCT in one location and wishes to know the treatment effect in another location without running additional RCTs. By using our proposal, they could get bounds on the estimated treatment effects in the second location directly. We validate this proposal on the GAIN dataset which includes data from independent RCTs conducted in several selected counties in California to evaluate the impact of welfare-to-work programs on an individual's future income. 
We focus on two counties: Los Angeles and Riverside. Suppose we had only run the RCT in Los Angeles and would like to extrapolate the results to Riverside. If there are no unmeasured effect modifiers that would affect both the outcome and location likelihood, we can simply use a weighted Hajek-style estimator. The goal is to get an uncertainty quantification of the extrapolated results in Riverside with varying degrees of how strong the unmeasured covariate shift is. Since the GAIN dataset includes the RCT data from Riverside, it allows us to validate the estimated bounds from our proposal against the ground truth treatment effect estimates from the RCT that had been conducted in Riverside.

\begin{figure}[t]
	\centering
	\begin{tabular}{cc}
		\includegraphics[width=0.5\columnwidth]{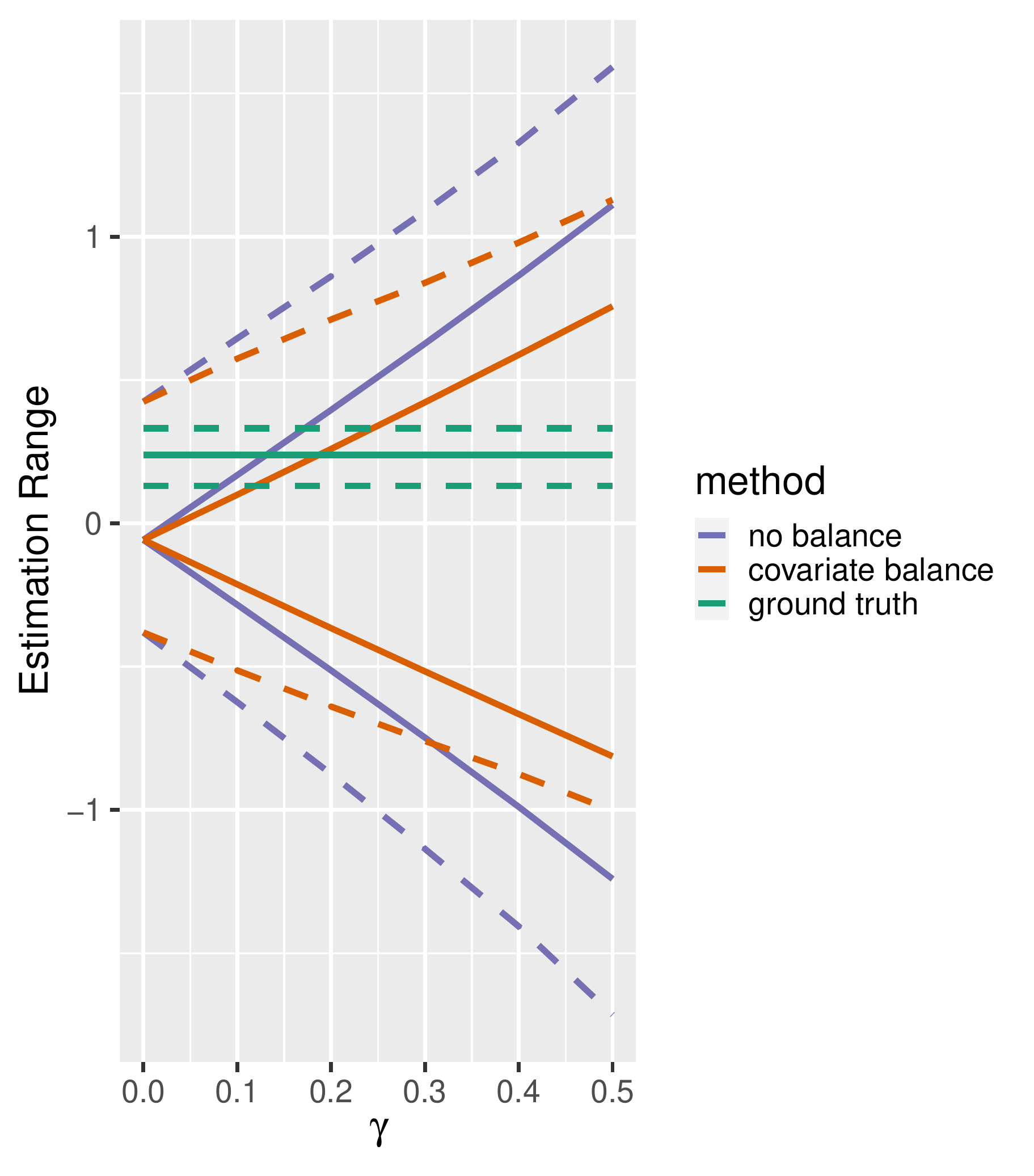}
		\includegraphics[width=0.5\columnwidth]{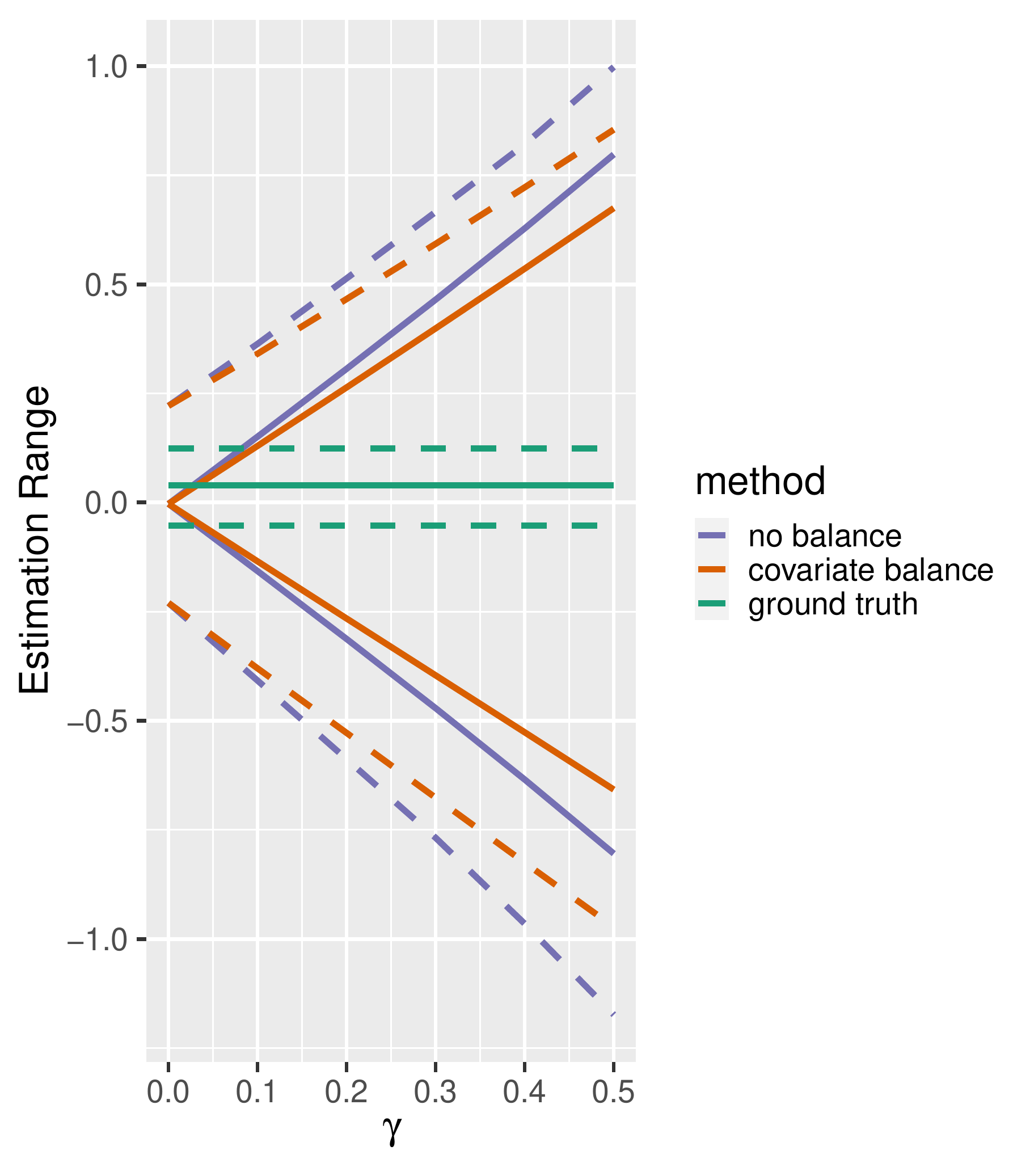} 
	\end{tabular}
	\vspace{-0.7\baselineskip}
	\caption{The left plot shows generalizing the RCT estimate from Los Angeles to Riverside, and the right plot shows generalizing the RCT estimate from Riverside to Los Angeles. The green line in both plots shows the difference-in-means estimate in the target location, which we denote by as the ``ground truth" treatment effect. The x-axis quantifies the assumed bound on unobserved distributional shift. The  red and purples  lines is our proposed estimator with and without covariate balancing respectively. We compute the 95\% confidence intervals via a percentile bootstrap with 1000 bootstrap samples (shown as dashed lines in the plots).
		\vspace{-\baselineskip}}
	\label{fig:gain}
\end{figure}

We run the proposed estimator with covariate balancing to generalize treatment effect bounds from Los Angeles to Riverside and vice versa (referred to as ``covariate balance" in Figure \ref{fig:gain}). For comparison, we also run the proposed estimator without leveraging covariate information (referred to as ``no balance" in Figure \ref{fig:gain}). We use the mean quarter income over a follow-up period of 9 years post experiment as the outcome. For both counties, we use a simple difference-in-means between the the treated and the control groups to compute their ground truth treatment estimates (referred to as ``ground truth" in the legend of Figure \ref{fig:gain}). 
Let $\gamma := \log \Gamma$. We quantify the strength of the unmeasured covariate shift by assuming different $\Gamma$ values in \eqref{eq:opt-no-bal-any-propensity} and \eqref{eq:bal-opt} for the ``no balance" and "covariate balance" approaches respectives.

The left plot in Figure \ref{fig:gain} shows generalizating the RCT estimate from Los Angeles to Riverside and the right plot shows the generalization results the other way. By varying $\gamma$ along the x-axis, we vary the assumed bound on the distributional shift of the unmeasured effect modifiers. We see that the ``covariate balance" approach meaningfully shortens the estimated bounds compared to the ``no balance" approach, and the bounds give coverage to the ground truth estimates as $\gamma$ increases. To estimate the variance, we generate 1000 bootstrap samples in both locations simultaneously to account for stochastic fluctuations in the data. 
The dashed lines show the 95\% confidence intervals using the percentile bootstrap, following \citet{zhao2019sensitivity}.



\section{Additional Simulations}
We consider the following setups adapted from the simulation study in \citet{yadlowsky2018bounds}. For some covariate distribution $P$,
\begin{align*}
	&X _k \sim P \ \ \ \textrm{for } k = 1,2,3,4, \  \  \   \   \   \\
	& S\sim Bernoulli\p{\frac{1-{\Gamma^*}^{-1} + (\Gamma^*-1) \exp(\alpha_0 + x^\top \mu)}{\Gamma^* - {\Gamma^*}^{-1} +  (\Gamma^* - {\Gamma^*}^{-1}) \exp(\alpha_0 + x^\top \mu)}}, \\
	&U' \sim \mathcal{N}(0, (1+0.5\sin(2.5X_1))^2), \ \ \ \ \ \  U = (2S-1) \abs{U'}\\
	& W \sim Bernoulli(0.5) \\
	& L \sim Bernoulli\Bigg\{ \frac{\exp(\alpha_0 + x^\top \mu + \log(\Gamma^*) (\mathbbm{1}_{U\geq0} - \mathbbm{1}_{U < 0}))}{1+\exp(\alpha_0 + x^\top \mu + \log(\Gamma^*) (\mathbbm{1}_{U\geq0} - \mathbbm{1}_{U < 0}))} \Bigg\} \\
	& \tau = X_1 + U + 4 \\
	&Y = (X+4)^\top \beta + U + W\tau + \sigma \epsilon, \ \ \ \ \ \epsilon \sim N(0,1)
\end{align*}
where $L=0,1$ denotes the location indicator. We consider the following two setups:

For Setup A, we let  $X_k \sim Beta(0.5, 0.5)$ for $k=1,2,3,4$ and $\mu = [2,2,-2,-2]$ such that $P(X\cond L)$ highly depends on the location $L$. We vary $\gamma \in [0,\,0.1,\,0.2, \,0,3,\,0.4,\,0.5]$ and let $\sigma = 3$ and $\alpha_0=0$. We let $\Gamma^* = exp(0.2)$ be the true sensitivity parameter, but we assume it's unknown to us.
For Setup B, we let $X \sim Uniform[0,1]^4$, $\Gamma^* = \exp(0.5)$, $\alpha_0 = -2, \sigma = 0.5$, and vary the sensitivity parameter $\gamma = \log{\Gamma} = 0, \,0.1, \,\ldots,\,0.7$ in the optimization procedure,  We let $\mu = [0.709, 0.438, 0.2, 0.767]$ in Setup B and let $\beta = [0.513, 0.045, 0.7, 0.646]$ in both setups.\footnote{Both parameters are taken from the simulation in \citet{yadlowsky2018bounds}}

\begin{figure}[t]
	\centering
	\begin{tabular}{cc}
		\includegraphics[width=0.5\columnwidth]{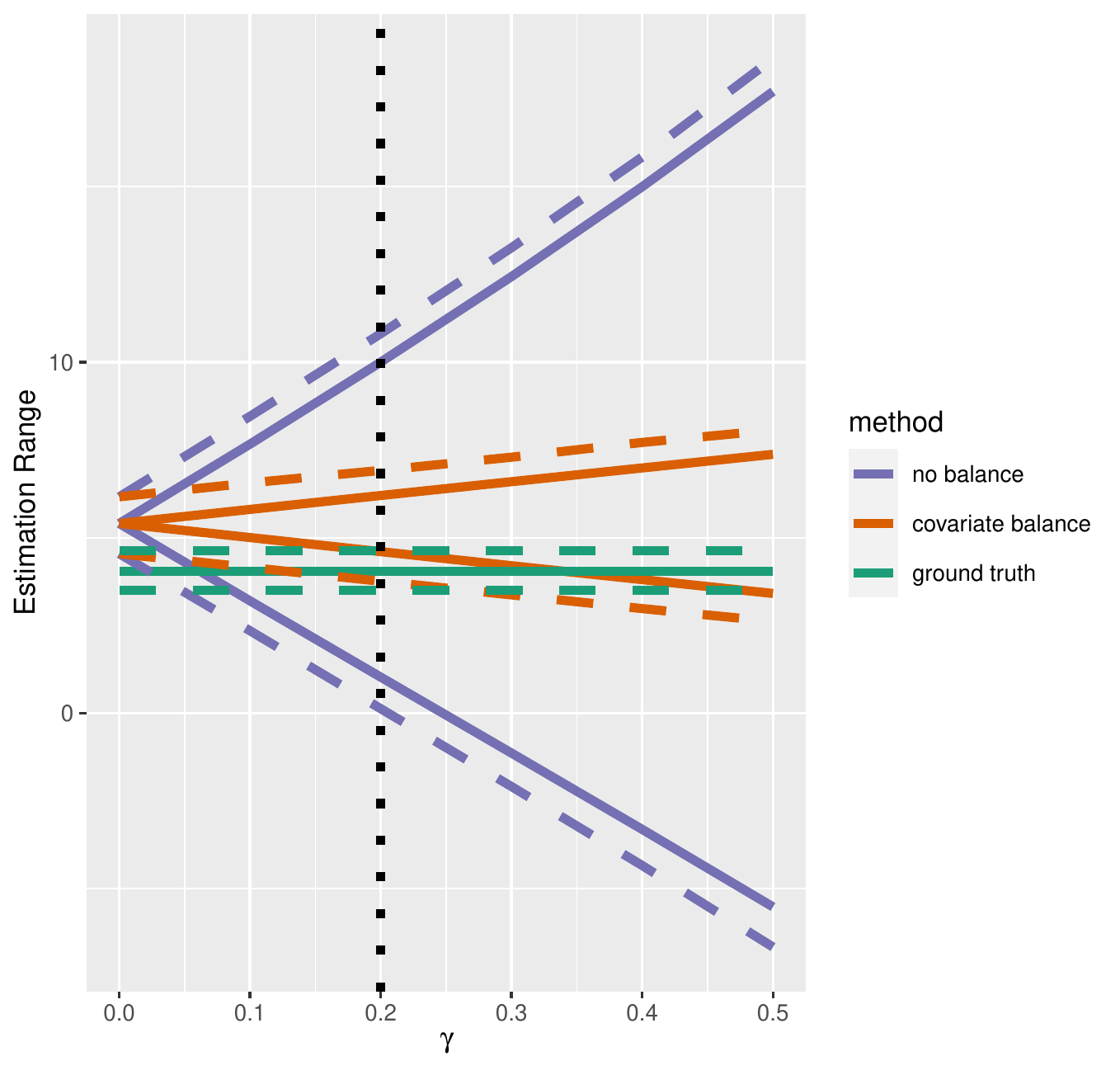}
		\includegraphics[width=0.5\columnwidth]{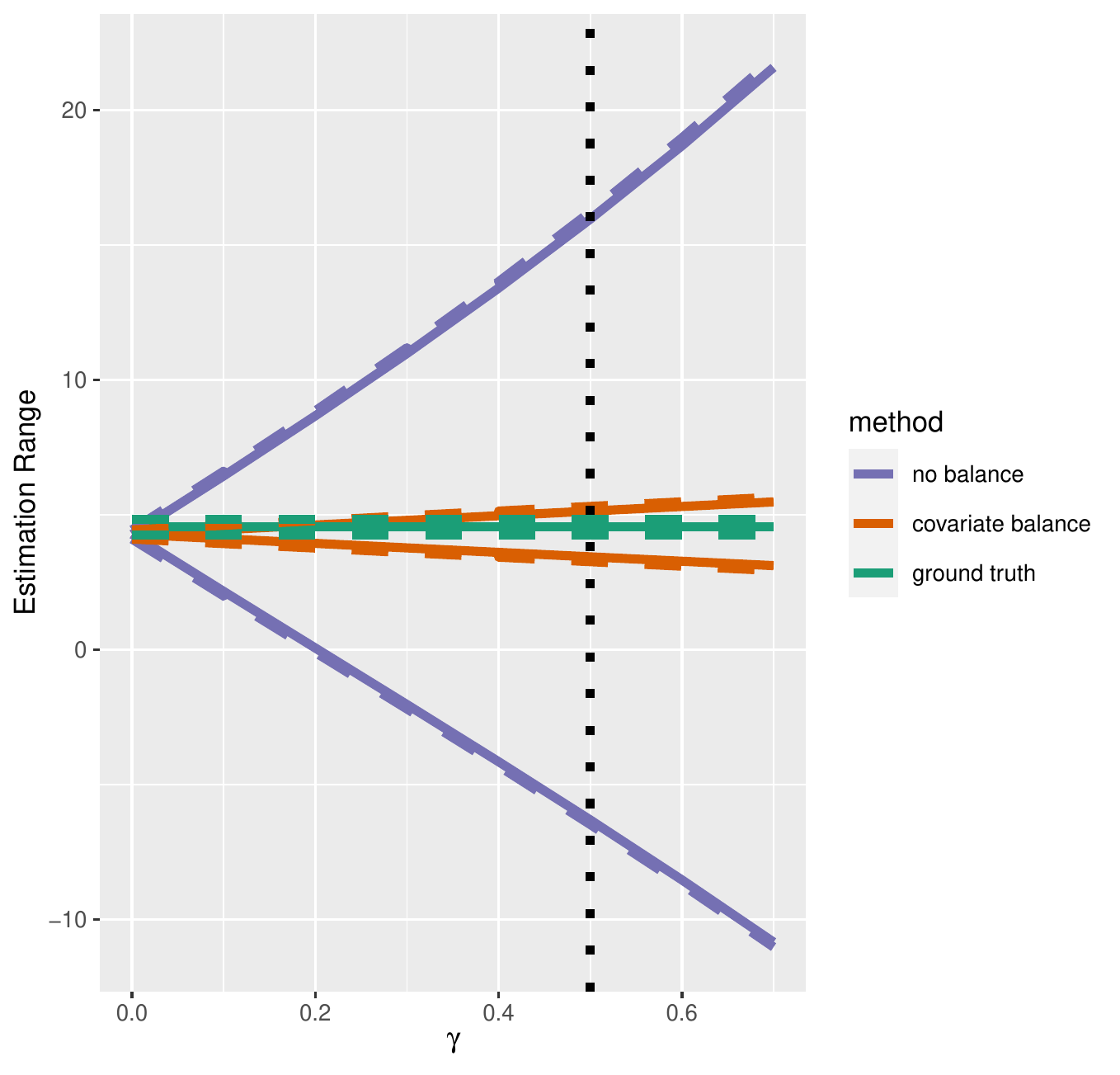} 
	\end{tabular}
	\vspace{-0.7\baselineskip}
	\caption{The two plots above show generalizing the RCT estimate in Setup A and B respectively for a randomly chosen single realization of the data generating distribution. The x-axis quantifies the assumed bound on unobserved distributional shift. We assume we have access to data in both locations, but not the ground truth treatment effects. The green line in both plots shows the difference-in-means \textit{estimate} of the ground truth in the target location. The red and purples lines are our proposed estimator with and without covariate balancing respectively. Given the single realization of the data generating distribution, we take 1000 bootstrap samples in both locations simultaneously, and compute the 95\% percentile bootstrap confidence intervals (shown as colored dashed lines in the plots). The vertical dotted black line shows the true sensitivity parameter $\gamma^* = \log(\Gamma^*) = 0.2$ in Setup A and $\gamma^* = \log(\Gamma^*) = 0.5$ in Setup B.
		\vspace{-\baselineskip}}
	\label{fig:gain-sim}
\end{figure}

 The form of $U$ is chosen such that Assumption \ref{assu:marginal-sensi-transport} holds with $\Gamma^*$, with $l(x) = \exp(\alpha_0 + x^\top \mu )/{\{1+\exp(\alpha_0 + x^\top \mu )\}}$, and $P(U=u\cond X=x, L=1) / P(U=u\cond X=x, L=0) = {\Gamma^*}^{\mathbbm{1}_{u>0} - \mathbbm{1}_{u < 0}}$.  and we let the size of the total combined population across the two locations to be 1000.\footnote{We note that for the purpose of this simulation setup, it is natural to define location as an additional random variable in the data generating process to ensure the ground truth sensitivity bound to fall within $[1/\Gamma^*, \Gamma^*]$.} 
 We use the \texttt{mosek} package for optimization, and we compare the percentile bootstrap confidence interval obtained through 1000 bootstrap samples among the difference-in-means estimator in location $L=1$ and the proposed estimator with covariate balancing. 
 We see that the covariate balancing approach signficantly shortens the estimation interval, while Setup A in Figure \ref{fig:gain-sim} also shows that our balancing estimator (in red) is not conservative as its confidence interval just covers the ground truth (in green) once the $\gamma$ parameter is increased to $\gamma^*=0.2$ in this case.

\newpage
\bibliographystyle{plainnat}
\bibliography{references}

\begin{appendix}
	\section{Proof of Proposition \ref{prop:1}}
	
	\begin{align*}
		\tau &:= \EE[1]{Y(1) - Y(0)} \\
		&=  \EE[1]{\EE[1]{Y(1) - Y(0) \cond X, U}} \\
		&= \EE[1]{\EE[0]{Y(1) - Y(0) \cond X, U}} \\
		&= \EE[0]{r(X,U)\EE[0]{Y(1) - Y(0) \cond X, U}} 
	\end{align*}
	where the last two equalities used Assumptions \ref{assu:relevant-select} and \ref{assu:support} respectively.
	

	\section{Proof of Theorem \ref{thm:consistency-1}}
\begin{proof}
	Assume $\Gamma = \Gamma^*$, let $\hat{r}_i(x, u) = \hat{r}_i(x) \Gamma^*$, then $\Norm{\hat{r}(x,u) - r(x,u)}_\infty \to_p 0$. By Proposition \ref{prop:1},
	\begin{align*}
		\frac{\sum_{i:L_i=0} W_iY_i \hat{r}_i(X_i, U_i)}{\sum_{i:L_i=0}W_i} \to_p \EE[1]{Y(1)}.
	\end{align*}
We can then show that 
\begin{align*}
	\frac{\sum_{i:L_i=0} W_iY_i \hat{r}_i(X_i, U_i)}{\sum_{i:L_i=0}W_i} - 	\frac{\sum_{i:L_i=0} (1-W_i)Y_i \hat{r}_i(X_i, U_i)}{\sum_{i:L_i=0}(1-W_i)} \to_p \tau.
\end{align*}
Since $\Gamma \geq \Gamma^*$, the conclusion then follows.

\end{proof}
	\section{Proof of Theorem \ref{thm:consistency}}
	\begin{proof}
		Recall $z_i^* = P_1(U_i\cond X_i)/P_0(U_i\cond X_i)$ is the oracle weight. 
		We note that by Proposition \ref{prop:z} and covariate balancing with the density ratio, taking $z_i=1$ is always a feasible solution to the optimization formulation \eqref{eq:bal-opt}. Thus, Slater's conditions hold. 
		Define 
		\begin{align*}
			\gamma_{i,n_0} = \frac{W_i(1-L_i) e^{\hat{g}_\phi(X_i)} }{\sum_{i: L_i=0} W_i } 
		\end{align*}
		Define $c_{n_1} =  \frac{\sum_{i: L_i=1} \phi(X_i)}{n_1 }$. Then we have
		\begin{align*}
			&\inf_\lambda \sum_{i: L_i=0} \gamma_{i,n_0} z_i^* Y_i - \lambda \p{\sum_{i: L_i=0} \gamma_{i,n_0}z_i^* \phi(X_i)  - c_{n_1}  }\\
			&\ \ \ \ \ \  \leq \hat{\mu}_1^+ := \inf_\lambda \sup_{z_1,\cdots, z_n} \sum_{i: L_i=0} \gamma_{i,n_0} z_i Y_i - \lambda \p{\sum_{i: L_i=0} \gamma_{i,n_0}z_i \phi(X_i)  - c_{n_1} }. 
		\end{align*}
		
		By Assumption \ref{assu:logistic}, $\Norm{\hat{g}_\phi(x) - g(x)}_\infty \to_p 0$ (see e.g. \citet{sugiyama2012density}). 
		By law of large numbers, we have as $n_1 \to \infty$, $$c_{n_1}  \to_p \EE[1]{\phi(X_i)},$$ and as $n_0 \to \infty$, by \eqref{eq:balance-x}, $$\sum_{i: L_i=0} \gamma_{i,n_0}z_i^* \phi(X_i)  \to_p \EE[1]{\phi(X_i)},$$ and
		$$\sum_{i: L_i=0} \gamma_{i,n_0} z_i^* Y_i  \to_p \mu_1$$
		where $\mu_1 = \EE[1]{Y(1)}$.
		
		 Conside a joint sequence of data generating setups for both locations with $\{n_0, n_1\}$ where $n_0, n_1\to \infty$, given $n_0/n_1 \to\eta$,  there must exist $N >0$ such that for all $n_0, n_1 \geq N$, $\eta/2 \leq n_0/n_1 \leq 2\eta$. From the derivation above, we know that for any $\delta>0$ and for any $\epsilon > 0$, there exists $N_1$ such that for all $n_1\geq N_1$, $P(\abs{c_{n_1} - \EE[1]{\phi(X_i)}} \geq \epsilon/2) \leq \delta/2$. Similarly,  there exists $N_0$ such that for all $n_0\geq N_0$, $P( \abs{\EE[1]{\phi(X_i)}  - \sum_{i: L_i=0} \gamma_{i,n_0}z_i^* \phi(X_i)}\geq \epsilon/2) \leq \delta/2$. Then, take $N^*\max\{N, N_0, N_1\}$, we have for $N_0, N_1 \geq N^*$, $P( \abs{c_{n_1} - \sum_{i: L_i=0} \gamma_{i,n_0}z_i^* \phi(X_i)}\geq \epsilon) \leq \delta$. 
		 
		 Thus, taking this joint sequence $\{n_0, n_1\}$ where $n_0, n_1\to\infty$ and $n_0/n_1 \to \eta$, we have
		 $$\sum_{i: L_i=0} \gamma_{i,n_0}z_i^* \phi(X_i)  - c_{n_1}  \to_p 0.$$
		 We then conclude that $\inf_\lambda \sum_{i: L_i=0} \gamma_{i,n_0} z_i^* Y_i - \lambda \p{\sum_{i: L_i=0} \gamma_{i,n_0}z_i^* \phi(X_i)  - c_n  } \to_p \mu_1$. 
		 
		 For any $\epsilon > 0$, we then have $\lim_{n_0, n_1\to \infty, n_0/n_1 \to \eta} P(\mu_1-\epsilon \leq \hat{\mu}_1^+) = 1$. Similarly, for any $\epsilon > 0$, we have $\lim_{n_0, n_1\to \infty, n_0/n_1 \to \eta} P(\mu_1+\epsilon \geq \hat{\mu}_1^-) = 1$, $\lim_{n_0, n_1\to \infty, n_0/n_1 \to \eta}  P(\mu_0 - \epsilon \leq \hat{\mu}_0^+) = 1$, and $\lim_{n_0, n_1\to \infty, n_0/n_1 \to \eta}  P(\mu_0 + \epsilon \geq \hat{\mu}_0^-) = 1$. We conclude that for any $\epsilon > 0$, $\lim_{n_0, n_1\to \infty, n_0/n_1 \to \eta}  P(\tau = \mu_1 - \mu_0 \in (\hat{\mu}_1^- - \hat{\mu}_0^+ -\epsilon, \hat{\mu}_1^+ - \hat{\mu}_0^- + \epsilon)) = 1$.

	\end{proof}

\section{An example to show how covariate balancing shortens estimation intervals}
To provide more intuition why covariate balancing shortens estimation intervals in Section \ref{sec:shorten}, we consider the simple context where the covariates $X$ are discrete and the link function $\phi$ is the identity function. Consider the following: 

Define
\begin{align*}
	\gamma_{i,n_0} = \frac{W_i(1-L_i) e^{\hat{g}_\phi(X_i)} }{\sum_{i: L_i=0} W_i } ,
\end{align*}
and define $c_{n_1} =  \frac{\sum_{i: L_i=1}  \phi(X_i)}{n_1 }$. Then we have
\begin{align*}
	\hat{\mu}_1^+ := \inf_{\substack{\lambda, \\h_1, \cdots, h_{n_0}, \\g_1, \cdots, g_{n_0}}} \sup_{z_1,\cdots, z_{n_0}}& \sum_{i: L_i=0} \gamma_{i,n_0} z_i Y_i + \lambda \p{\sum_{i: L_i=0} \gamma_{i,n_0}z_i \phi(X_i)  - c_{n_1} } + \sum_{i:L_i=0 } h_i (\Gamma - z_i) \\
	&+ \sum_{i: L_i=0} g_i(z_i - \Gamma^{-1}).
\end{align*}
Now let $\phi$ be identity, and let $X$ be binary, then
\begin{align*}
	\hat{\mu}_1^+ := \inf_{\substack{\lambda, \\h_1, \cdots, h_{n_0}, \\g_1, \cdots, g_{n_0}}} \sup_{z_1,\cdots, z_{n_0}} &\sum_{i: L_i=0} \gamma_{i,n} z_i Y_i + \lambda \p{\sum_{i: L_i=0, X_i=1} \gamma_{i,n}z_i  - c_{n_1} }\\
	& + \sum_{i: L_i=0} h_i (\Gamma - z_i) + \sum_{i: L_i=0} g_i(z_i - \Gamma^{-1}).
\end{align*}

Compared to the approach without covariate balancing,
\begin{align*}
	\hat{\mu}_{1, \textrm{no bal}}^+ := \inf_{\substack{h_1, \cdots, h_{n_0}, \\g_1, \cdots, g_{n_0}}} \sup_{z_1,\cdots, z_{n_0}} \sum_{i: L_i=0} \gamma_{i,n} z_i Y_i + \sum_{i: L_i=0} h_i (\Gamma - z_i) + \sum_{i: L_i=0} g_i(z_i - \Gamma^{-1}).
\end{align*}

Essentially we want to maximize $ \sum_{i: L_i=0} \gamma_{i,n} z_i Y_i$, but with covariate balancing, we limit the sum of certain weights such that $\sum_{i: L_i=0, X_i=1} \gamma_{i,n}z_i = c_n$. 

	\section{Proof of Theorem \ref{thm:consistency-approx}}
	\begin{proof}
		Recall $z_i^* = P_1(U_i\cond X_i)/P_0(U_i\cond X_i)$ is the oracle weight. Given the ground truth sensitivity parameter $\Gamma^* \leq \Gamma$ and  $M \geq e^{g(\cdot)} / e^{g_\phi(\cdot)}$, then $\Gamma^{-1}  M^{-1} \leq z_i^* \leq \Gamma M$ for all $1\leq i \leq n_0$ by Proposition \ref{prop:z-misspe}.
		We note that by covariate balancing with density ratios, taking $z_i=1$ is always a feasible solution to the optimization formulation \eqref{eq:bal-opt}. Thus, Slater's conditions hold. 
		Define 
		\begin{align*}
			\gamma_{i,n_0} = \frac{W_i(1-L_i) e^{\hat{g}_\phi(X_i)}}{\sum_{i: L_i=0} W_i } 
		\end{align*}
		Define $c_{n_1} =  \frac{\sum_{i: L_i=1} \phi(X_i)}{n_1}$. Then we have
		\begin{align*}
			&\inf_\lambda \sum_{i: L_i=0} \gamma_{i,n_0} z_i^* Y_i - \lambda \p{\sum_{i: L_i=0} \gamma_{i,n_0}z_i^* \phi(X_i)  - c_{n_1}  }\\
			&\ \ \ \ \ \  \leq \hat{\mu}_{\phi,1}^+ := \inf_\lambda \sup_{z_1,\cdots, z_{n_0}} \sum_{i: L_i=0} \gamma_{i,n_0} z_i Y_i - \lambda \p{\sum_{i: L_i=0} \gamma_{i,n_0}z_i \phi(X_i)  - c_{n_1} }. 
		\end{align*}
		
		By law of large numbers, as $ n_1\to \infty$, 
		$$c_{n_1}  \to\EE[1]{\phi(X_i)},$$
		and as $n_0 \to \infty$,
		$$\sum_{i: L_i=0} \gamma_{i,n_0}z_i^* \phi(X_i) \to \EE[1]{\phi(X_i)}$$
		by \eqref{eq:balance-x}, and
		$$\sum_{i=1: L_i=0} \gamma_{i,n_0} z_i^* Y_i  \to_p \mu_1$$
		 where $\mu_1 = \EE[1]{Y(1)}$.
		
		As we take both $n_0, n_1\to\infty$ and as $n_0/n_1\to \eta$, following a similar argument used in the proof of Theorem \ref{thm:consistency},
		$$\sum_{i: L_i=0} \gamma_{i,n_0}z_i^* \phi(X_i)  - c_{n_1}  \to_p 0.$$
		Thus, $\inf_\lambda \sum_{i: L_i=0} \gamma_{i,n_0} z_i^* Y_i - \lambda \p{\sum_{i: L_i=0} \gamma_{i,n_0}z_i^* \phi(X_i)  - c_{n_1}  } \to_p \mu_1$. 
		
		For any $\epsilon>0$, $\lim_{n_0, n_1\to \infty} P(\mu_1 \leq \hat{\mu}_{\phi,1}^+ + \epsilon) = 1$. Similarly, for any $\epsilon > 0$, we have $\lim_{n_0, n_1\to \infty} P(\mu_1 \geq \hat{\mu}_{\phi,1}^- - \epsilon) = 1$, $\lim_{n_0, n_1\to \infty} P(\mu_0 \leq \hat{\mu}_{\phi,0}^+ + \epsilon) = 1$, and $\lim_{n_0, n_1\to \infty} P(\mu_0 \geq \hat{\mu}_{\phi,0}^- - \epsilon) = 1$. We conclude that $\lim_{n_0, n_1\to \infty} P(\tau = \mu_1 - \mu_0 \in (\hat{\mu}_{\phi,1}^- - \hat{\mu}_{\phi,0}^+ - \epsilon, \hat{\mu}_{\phi,1}^+ - \hat{\mu}_{\phi,0}^- + \epsilon) ) = 1$
	\end{proof}

\end{appendix}

\end{document}